\documentclass[12pt]{article}

\usepackage{graphicx,amsmath,amssymb,url}
\usepackage{slashed}    

\newlength{\dinwidth}
\newlength{\dinmargin}
\def\lapproxeq{\lower .7ex\hbox{$\;\stackrel{\textstyle
<}{\sim}\;$}}
\def\gapproxeq{\lower .7ex\hbox{$\;\stackrel{\textstyle
>}{\sim}\;$}}
\def\gtrsim{\lower .7ex\hbox{$\;\stackrel{\textstyle
>}{\sim}\;$}}
\def\lesim{\lower .7ex\hbox{$\;\stackrel{\textstyle
<}{\sim}\;$}}

\def\be{\begin{equation}}
\def\ee{\end{equation}}
\def\bea{\begin{eqnarray}}
\def\eea{\end{eqnarray}}

\def\J{J/\psi}

\def\MSbar{{\overline {\rm MS}}}
\newcommand{\eq}[1]{(\ref{eq:#1})}

\begin{document}

\titlepage

\begin{flushright}
MPP-2016-303\\
IPPP/16/87\\

LTH 1102\\

\today\\

\end{flushright}

\vspace*{2cm}

\begin{center}

{\Large \bf The exclusive $J/\psi$ process at the LHC\\
\vspace{0.5cm}

 tamed to probe the low $x$ gluon}

\vspace*{1cm} {\sc S.P. Jones}$^{a}$, {\sc A.D. Martin}$^b$,  {\sc
  M.G. Ryskin}$^{b, c}$ and {\sc T. Teubner}$^{d}$ \\

\vspace*{0.5cm}
$^a$ {\em Max-Planck-Institute for Physics, F\"{o}hringer Ring 6, 80805 M\"{u}nchen, Germany}\\
$^b$ {\em Institute for Particle Physics
  Phenomenology, Durham University, Durham DH1 3LE, U.K.}\\

$^c$ {\em Petersburg Nuclear Physics Institute, NRC Kurchatov Institute, Gatchina,
St.~Petersburg, 188300, Russia} \\
$^d$ {\em Department of Mathematical Sciences,
University of Liverpool, Liverpool L69 3BX, U.K.}\\
 \end{center}

\vspace*{1cm}

\begin{abstract}

The perturbative QCD expansion for $\J$ photoproduction appears to be unstable: the NLO correction is large (and of opposite sign) to the LO contribution. Moreover, the predictions are very sensitive to the choice of factorization and renormalization scales.  Here we show that perturbative stability is greatly improved by imposing a $`Q_0$ cut' on the NLO coefficient functions; a cut which is required to avoid double counting. $Q_0$ is the input scale used in the parton DGLAP evolution. This result opens the possibility of high precision exclusive $\J$ data in the forward direction at the LHC being able to determine the low $x$ gluon distribution at low scales.

\end{abstract}

\vspace*{0.5cm}

\section{Introduction}

\begin{figure} [h]
\begin{center}
\includegraphics[width=0.4\textwidth]{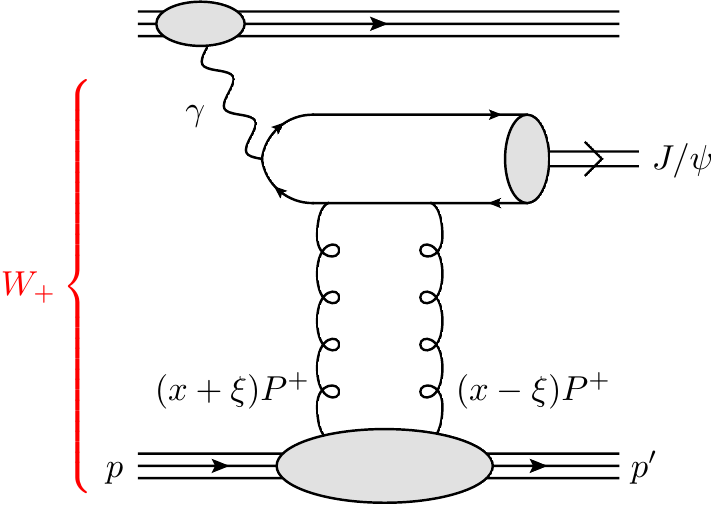} 
\qquad
\includegraphics[width=0.4\textwidth]{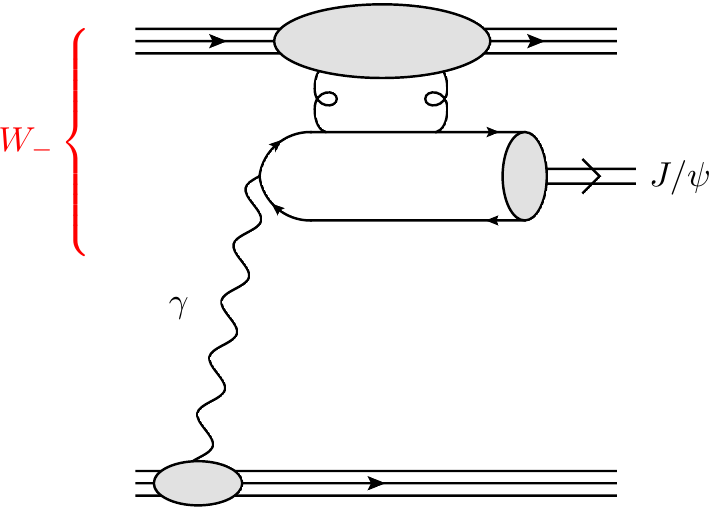}
 \caption{ $d\sigma(pp\to p+\J+p)/dy$ driven by the subprocess $\gamma p\to \J+p$ at two different $\gamma p$ centre-of-mass energies, $W_\pm$. }
\label{fig:subproc}
\end{center}
\end{figure}

It would be valuable to be able to constrain the gluon  parton distribution function (PDF) at low $x$ using $\J$ photoproduction data measured at HERA and at the LHC, via exclusive $pp \to p+\J+p$ events, especially events in the forward region measured by the LHCb collaboration. Indeed, for LHCb kinematics at 13 TeV we can reach down to $x\simeq 3\times 10^{-6}$. 
Exclusive $\J$ production is driven by the subprocess $\gamma^*p\to \J +p$, see Fig.~\ref{fig:subproc}.
Unfortunately, it turns out that the NLO corrections calculated in the conventional $\MSbar$ collinear approach are found to be very large and to depend strongly on the choice of factorization and renormalization scales \cite{ISSK,IPSW,JMRT}. Indeed, for an `optimum' choice of scales it is found that the NLO correction has the opposite sign to the LO contribution and even changes the sign of the whole amplitude, see the continuous curves in Fig.~\ref{fig:f1}. Thus one may doubt the convergence of the whole perturbation series.

\subsection{Optimum scale \label{sec:opt} }
What do we mean by the `optimum' scale?
It was shown in Ref.~\cite{JMRT} that it is possible to find a scale (namely $\mu_F=m_c$) which resums all the double logarithmic corrections enhanced by large values of ${\rm ln}(1/\xi)$ into the gluon and quark PDFs, where $\xi$ is the skewedness parameter of the Generalised Parton Distributions (GPDs) describing the proton-gluon (and proton-quark) vertices. 
That is, it is possible to take the $(\alpha_S{\rm ln}(1/\xi){\rm ln}(\mu_F^2))$ term from the NLO gluon (and quark) coefficient functions and to move it to the LO GPDs.  This allows a resummation of all the double log  $(\alpha_S{\rm ln}(1/\xi){\rm ln}(\mu_F^2))^n$ terms in the LO contribution by choosing the factorization scale to be $\mu_F=m_c$.
The details are given in Ref.~\cite{JMRT}, see also Ref.~\cite{DY}.

The result is that the $\gamma p\to \J+p$ amplitudes are schematically of the form
\be
A(\mu_f)~=~C^{\rm LO} \otimes {\rm GPD}(\mu_F)~+~C^{\rm NLO}_{\rm rem}(\mu_F)\otimes{\rm GPD}(\mu_f),
\ee
where the GPD can be related to the conventional PDF via the Shuvaev transform for $\xi<|x|\ll 1$ \cite{Shuv}.
With the choice $\mu_F=m_c$ there is a smaller remaining term in the NLO coefficent funcions, and so the residual dependence on the scale $\mu_f$ is reduced.

\begin{figure} [t]
\begin{center}
\includegraphics[width=0.4\textwidth,angle=-90]{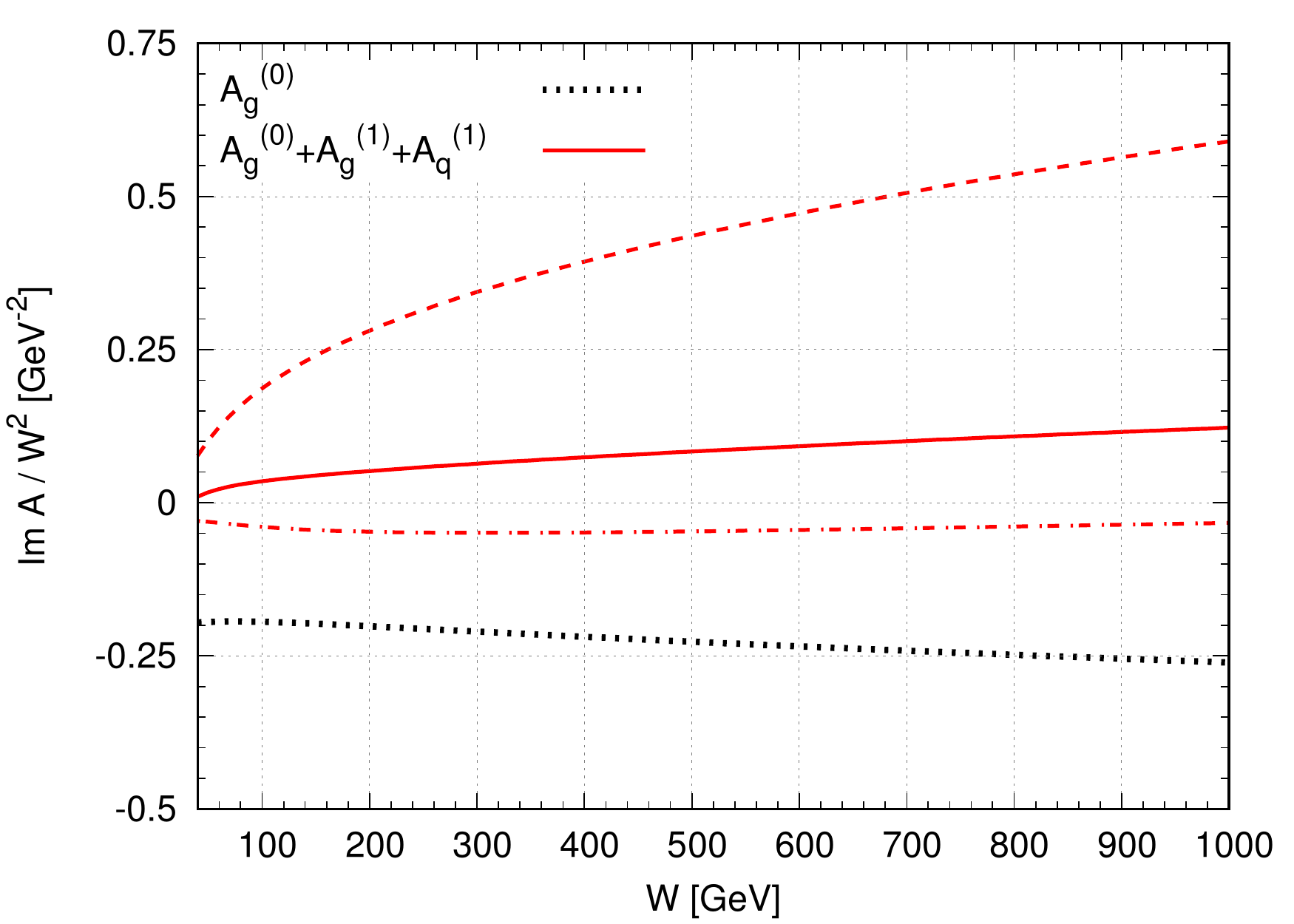} 
\caption{\sf 
The dotted and continuous curves are the LO and NLO predictions, respectively, of $\mathrm{Im} A/W^2$ for the $\gamma p \rightarrow \J + p$ amplitude, $A$,
as a function of the $\gamma p$ centre-of-mass energy $W$, 
obtained 
using CTEQ6.6 partons~\cite{Nadolsky:2008zw} (with input $Q_0=1.3$ GeV) for the optimal scale choice
$\mu_F = \mu_R =m_c$. The top
three curves correspond to the NLO prediction for various values of the residual factorization scale $\mu_f$, namely: $\mu_f^2 = ~2m_c^2, ~m_c^2, ~Q_0^2$ respectively where $m^2_c\equiv M^2_\psi /4=2.4$ GeV$^2$.}
\label{fig:f1}
\end{center}
\end{figure}

Unfortunately, even after this, the NLO corrections, and their variations with scale, although reduced, are still unacceptably large, as shown in Fig.~\ref{fig:f1}.  The dashed and dot-dashed curves correspond to NLO predictions for two different values of the residual scale $\mu_f$: namely $\mu_f^2=$ 4.8 and 1.7 GeV$^2$ respectively, while the continuous curves correspond to the `optimum' scale choice $\mu_F^2=\mu^2_R=m_c^2=M_\psi^2/4 =2.4 $ GeV$^2$.~\footnote{Recall that the choice $m_c=M_\psi/2$ effectively accounts for the relativistic corrections to the $\J$ wave function, see \cite{ Hoodbhoy,Noc}.}  The choice $\mu_R=\mu_F$ is justified in subsection \ref{sec:scales}.

\subsection{Double counting}

\begin{figure} [t]
\begin{center}
\includegraphics[width=0.4\textwidth]{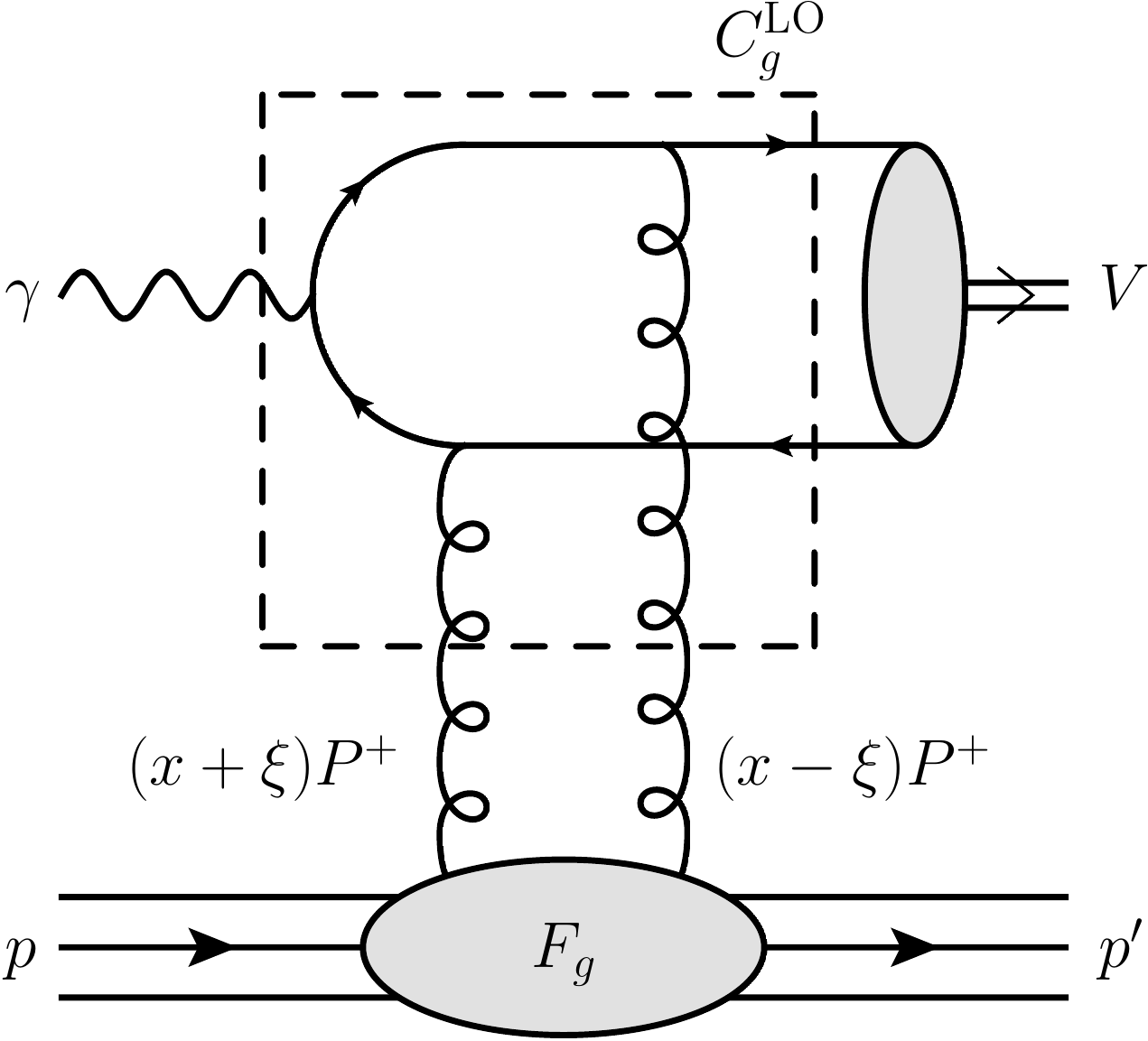}
\qquad
\includegraphics[width=0.4\textwidth]{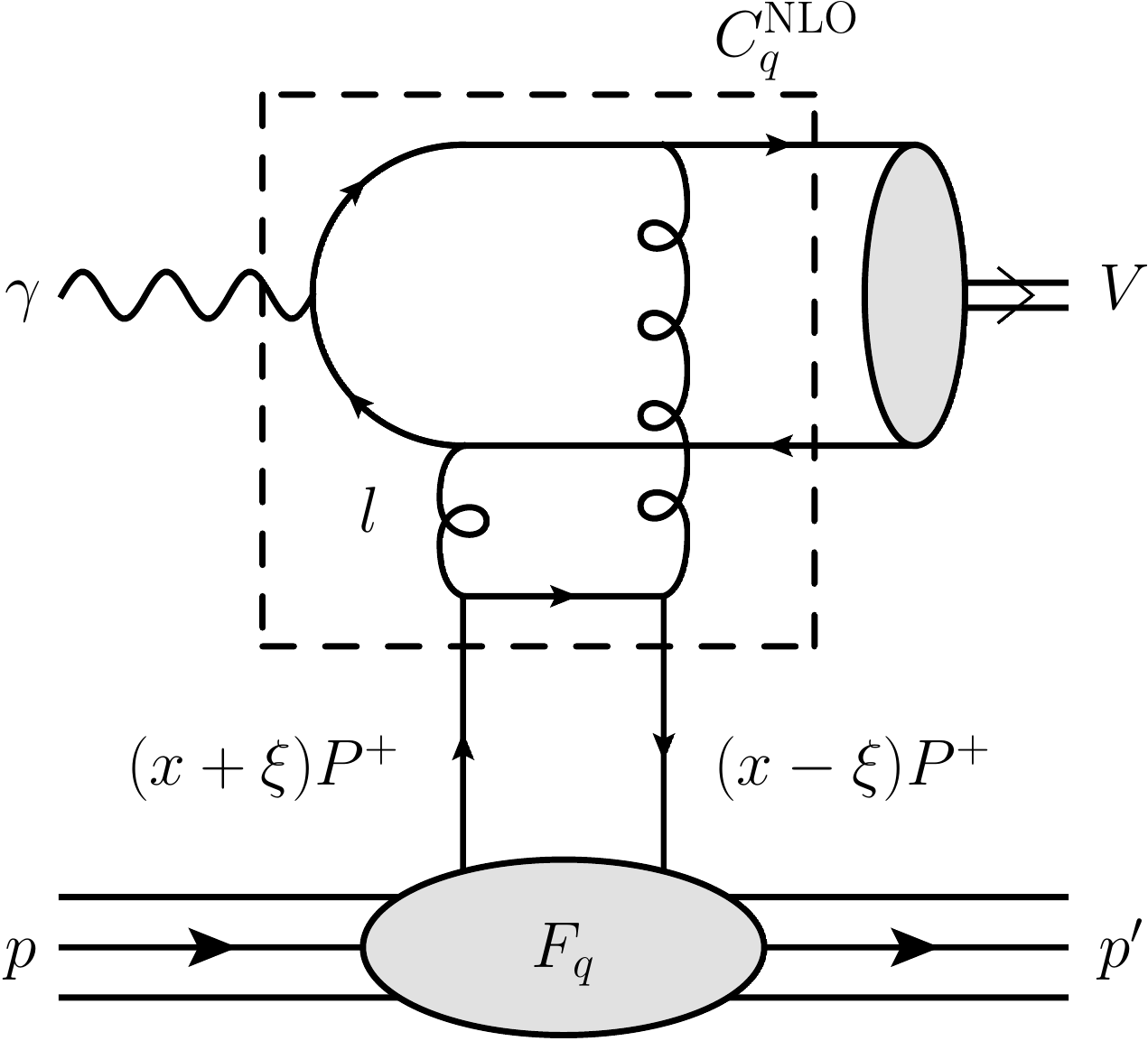}
\caption{\sf (a) LO contribution to  $\gamma p \to V +p$. (b) NLO quark contribution. For
  these graphs all permutations of the parton lines and couplings of
  the gluon lines to the heavy-quark pair are to be understood. Here
  $P\equiv (p+p^\prime)/2$ and $l$ is the loop momentum.
}
\label{fig:f2}
\end{center}
\end{figure}

So for the QCD prediction to be useful we should search for 
 some other sizeable physical contribution to the NLO correction.
Here we consider a power correction which may further reduce the NLO correction and, moreover, may reduce the sensitivity to the choice of scale. The correction is ${\cal O}(Q_0^2/M^2_\psi)$ where $Q_0$ denotes the input scale in the parton evolution.  It turns out to be important for the relatively light charm quark, $m_c \simeq M_\psi /2$. Let us explain the origin of this `$Q_0$ correction'.  We begin with the collinear factorization approach at LO. Here, we never consider parton distributions at low virtualities, that is for $Q^2<Q_0^2$.  We start the PDF evolution from some phenomenological PDF input at $Q^2=Q_0^2$. In other words, the contribution from  $|l^2|<Q^2_0$ of Fig.~\ref{fig:f2}(b) (which can be considered as the LO diagram, Fig.~\ref{fig:f2}(a), supplemented by one step of DGLAP evolution from quark to gluon, $P_{gq}$)  is already included in the input gluon GPD at $Q_0$.
  That is, to avoid double counting, we must exclude from the NLO diagram the contribution coming from virtualities less than $Q_0^2$. At large scales, $Q^2\gg Q^2_0$ this double-counting correction will give small power suppressed terms of ${\cal O}(Q_0^2/Q^2)$, since there is no infrared divergence in the corresponding integrals.  On the other hand, with $Q_0 \sim 1$ GeV and $\mu_F=m_c~ (\sim M_\psi /2$) a correction of ${\cal O}(Q^2_0/m^2_c)$ may be crucial.

In the present paper we re-calculate the NLO contribution for $\J$ photoproduction excluding the contribution coming from the low virtuality domain $(<Q^2_0)$.  We find that for $\J$ this procedure substantially reduces the resulting NLO contribution and, moreover, reduces the scale dependence of the predictions. It indicates the convergence of the perturbative series. 

An outline of the procedure is given in \cite{Diff2016}, where also the NLO description of exclusive $\J$ production in the $k_T$ factorization and collinear factorization schemes are compared. 

\section{Avoiding double counting of the low $Q^2$ contribution}

\subsection{The NLO quark contribution}

We start with the NLO quark contribution to the $\gamma p \to \J +p$ process. The corresponding Feynman diagrams are that of Fig.~\ref{fig:f2}(b) together with the diagram where both gluons couple to the same heavy quark line. Here we will use the non-relativistic approximation for the $\J$ wave function.  Since the momentum fractions $(x+\xi)$ and $(x-\xi)$ carried by the left and right quarks are different we have to use the skewed (generalized) parton distribution (GPD), $F_q(x,\xi,Q^2)$. The skewedness parameter $\xi=M_\psi^2/(2W^2-M_\psi^2)$, where $W$ is the $\gamma p$ energy.
We see that the upper part of diagram Fig.~\ref{fig:f2}(b) is the same as the diagram for the LO gluon Fig.~\ref{fig:f2}(a) contribution. For the LO contribution the integral over the gluon virtuality $|l^2|$ starts from the input scale $Q^2_0$, while all the contributions from low virtualities $|l^2|<Q^2_0$ are collected in the input gluon GPD, $F_g(x,\xi,Q_0^2)$.  Note that this input distribution already includes that part of the quark contribution of Fig.~\ref{fig:f2}(b) coming from $|l^2|<Q^2_0$. Thus to avoid double counting when computing the NLO quark coefficient function, $C_q^{\rm NLO}$, of Fig.~\ref{fig:f2}(b) we have to include the theta function $\Theta(|l^2|>Q^2_0)$ in the integration over $l^2$.  Depending on the ratio $Q^2_0/m^2_c=4Q^2_0/M^2_\psi$ this can be a significant correction.
The corresponding integral has no infrared or ultraviolet divergence  and can be calculated in $D=4$ dimensions.

Actually, the calculation is performed in the physical scheme (with $D=4)$. On the other hand, parton distributions are usually presented in the $\MSbar$ factorization scheme where dimensional regularization is used. The problem is that when we  calculate the coefficient function in $D=4+2\epsilon$ we have finite contributions of $\epsilon/\epsilon$ origin. Formally these $\epsilon/\epsilon$ terms come from unphysically large distances  $\propto {\cal O}(1/\epsilon)$. In fact, these $\epsilon/\epsilon$ terms are compensated by a corresponding re-definition of the PDFs.  In order to retain the $\epsilon/\epsilon$ terms and to use the $\MSbar$ scheme we do not calculate diagram \ref{fig:f2}(b) in $D=4$ dimensions for $|l^2|>Q^2_0$, but instead calculate the part corresponding to small $|l^2|<Q^2_0$. We consider this part as the correction which should be subtracted from the known NLO $\MSbar$ coefficient function \cite{ISSK,SJonesThesis}. Recall that after the subtraction of the contribution generated by the last step of the LO evolution, $P^{\rm LO}\otimes C^{\rm LO}$, there is no infrared divergence and the subtracted part of $C^{\rm NLO}$ coming from $|l^2|<Q_0^2$ does not contain $\epsilon/\epsilon$ terms.

\subsection{The NLO gluon contribution}

The NLO `$Q_0$ corrections' for the gluon coefficient function are more complicated.  Besides the ladder-type diagrams analogous to Fig.~\ref{fig:f2}(b), but with the light quark line replaced by a gluon line, there are other diagrams which have a structure similar to vertex corrections, see \cite{ISSK,SJonesThesis}.  However the `dangerous' contribution is again from the ladder-type diagrams, where to avoid double counting we have to exclude the $|l^2|<Q^2_0$ domain whose contribution is already included in the LO term using the input gluon GPD, $F_g(x,\xi,Q^2_0)$. Qualitatively this is exactly the same calculation as that for the NLO quark. The only difference is that the lower line in the diagrams of Fig.~\ref{fig:f4} is now replaced by a gluon line and the lower part of the diagram
is now given by the product of two three-gluon vertices averaged over the incoming gluon transverse polarizations.
 The notation is identical to that for the quark contribution. Both the quark- 
and the gluon-induced contributions are determined as described in the Appendix. They involve the calculation of the diagrams of Fig.~\ref{fig:f4} (given in the Appendix), and the analogous diagrams for the gluon-induced contribution.

\section{Results}

\begin{figure} [t]
\begin{center}
\includegraphics[width=0.4\textwidth,angle=-90]{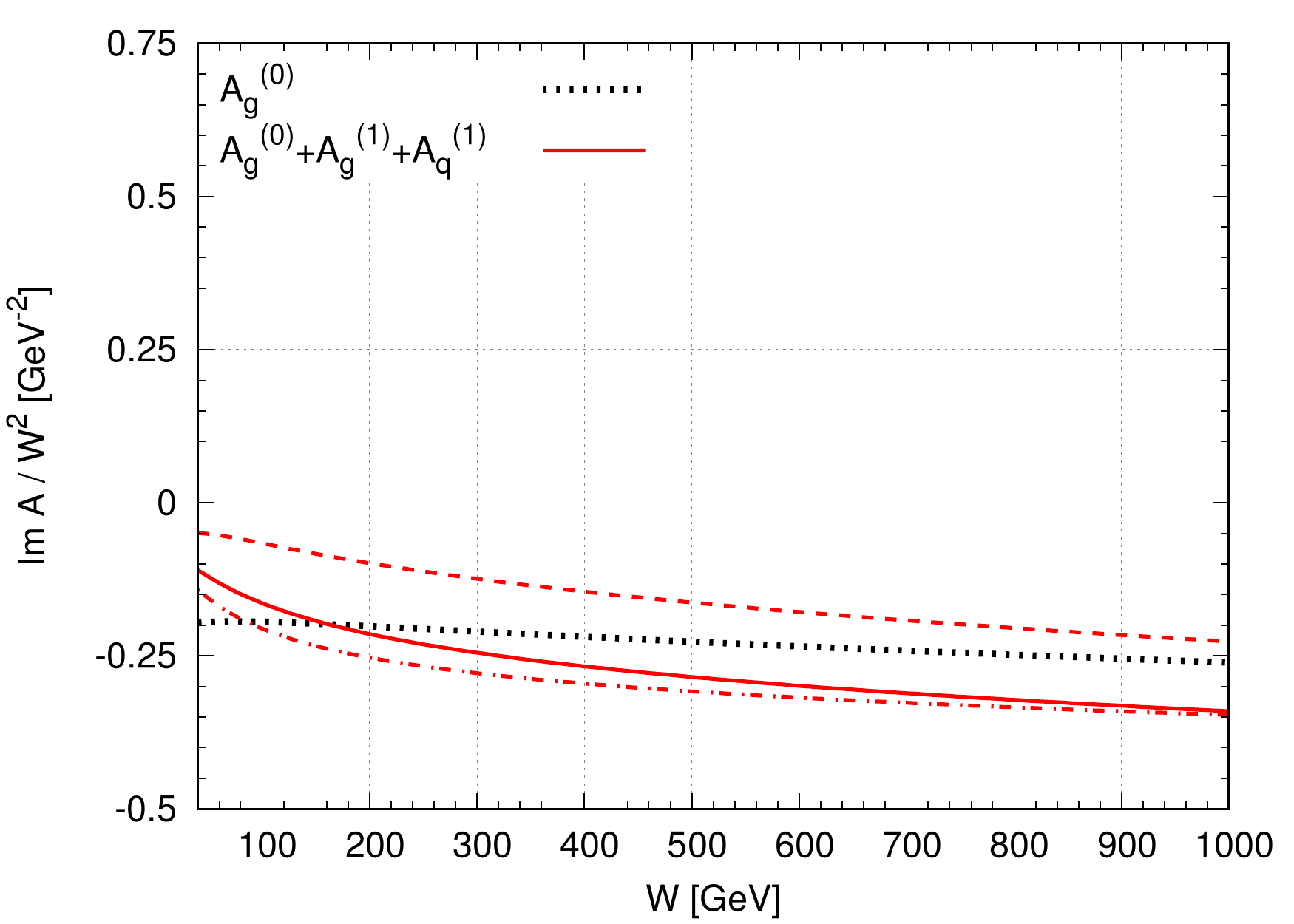}
\caption{\sf The predictions for the LO and NLO contributions to the imaginary part of the $\J$ photoproduction amplitude calculated exactly as in Fig.~\ref{fig:f1} except that now the $Q_0$ cut is imposed.}
\label{fig:f5}
\end{center}
\end{figure}

Fig.~\ref{fig:f5} shows the LO and NLO contributions to the imaginary part of the $\J$ photoproduction amplitude when the $Q_0$ cut in the NLO contribution is taken into account.  It should be compared to Fig.~\ref{fig:f1} which had exactly the same scale choices, but without the $Q_0$ cut imposed. The improvement in going from Fig.~\ref{fig:f1} to Fig.~\ref{fig:f5} is dramatic.
First, the NLO contribution is now much smaller than the LO contribution.  Second, the scale variation is much smaller.  The continuous curves in Figs.~\ref{fig:f1} and~\ref{fig:f5} show the LO and NLO comparison for the choice of scales $\mu_F=\mu_R=m_c\equiv M_\psi/2$, which we had previously argued was optimal \cite{JMRT}.
The stability achieved by imposing the $Q_0$ cut means that $J/\psi$ photoproduction ($\gamma p \to J/\psi ~p$) data and LHC exclusive $J/\psi$ ($pp \to p+J/\psi+p$) data can now be included in the global parton analyses.

\subsection{The choice of scales \label{sec:scales}}
Let us discuss the above scale choices in more detail. By choosing the `optimal' factorization scale $\mu_F=m_c$ we resum all the higher-order double-logarithmic corrections $(\alpha_s\ln(1/\xi)\ln\mu^2_F)^n$ (enhanced at high energies by the large value of $\ln(1/\xi)$) into the gluon generalized parton distribution (gluon GPD)~\cite{JMRT}. 

The renormalization scale is taken to be $\mu_R=\mu_F$.
The arguments are as follows. First, this corresponds to the BLM prescription~\cite{BLM};  such a choice eliminates from the NLO terms the contribution proportional to $\beta_0$ (i.e. the term $\beta_0\ln(\mu^2_R/\mu^2_F)$ in eq.~(3.95) of \cite{ISSK}). Second, following the discussion in~\cite{LHKR} for the analogous QED case, we note that the new quark loop insertion into the gluon propagator appears twice in the calculation. The part with scales $\mu <\mu_F$ is generated by the virtual component ($\propto \delta(1-z)$) of the LO splitting during DGLAP evolution, while the part with scales $\mu>\mu_R$ accounts for the running $\alpha_s$ behaviour obtained after the regularization of the ultraviolet divergence. In order not to miss some contribution and/or to avoid double counting we take the renormalization scale equal to the factorization scale, $\mu_R=\mu_F$.

\subsection{Discussion of the results}

Note that in the present paper we have calculated the imaginary part of the $\gamma p \to \J~p$ amplitude.
The real part of the amplitude can be restored via dispersion relations assuming positive signature, as in eq.~(5) of Ref.~\cite{JMRT1311}.  Recall that we obtain the necessary GPDs from the CTEQ6.6 parton set \cite{Nadolsky:2008zw} using the Shuvaev transform \cite{Shuv}. We use a relatively old parton set \cite{Nadolsky:2008zw} in which the low $x$ gluons are forced to be positive so as to make a meaningful comparison with our earlier work.  The goal of this paper is not to make a quantitative description of the data, but to demonstrate that we can achieve stability of the perturbative QCD description of relatively low scale $\J$ production by imposing the $Q_0$ cut. We have shown this is a power correction -- a correction which is needed to avoid double counting. This will allow future high precision exclusive $\J$ production data obtained at the LHC to be incorporated in global parton analyses.

The general procedure to include the HERA $\gamma p \to \J~p$ data and, in particular, the LHCb data for exclusive $\J$ production, $pp \to p+\J+p$, in a global analysis follows that used to produce Fig.~4 of Ref.~\cite{JMRT1311}. These processes are driven by the gluon PDF and the LHCb data probe the gluon at very low values of $x$.
However, in Ref.~\cite{JMRT1311} we {\it approximated} the NLO corrections to the coefficient functions by accounting for the explicit $l_\perp$ integration in the last step of the interaction. Moreover, we just fitted the $\J$ data
and used a parametric form for the gluon which approximated its $x$ and $Q^2$ dependence.  So the analysis of Ref.~\cite{JMRT1311} was quite simplified, although very informative; see, for example, Fig.~5 of \cite{JMRT1311} which compared the resulting gluon PDF with those of different global analyses\footnote{Recall, however, that strictly speaking the global analyses use the $\overline{\rm MS}$ collinear factorization scheme whereas $k_T$ factorization uses the physicsl scheme, see, for example, \cite{Diff2016}.}. 

The present paper, on the other hand, retains collinear factorization and calculates the complete NLO contribution. We may expect the high $\gamma p$ energy, $W$, data points in the updated version of Fig. 4 of Ref.~\cite{JMRT1311} to require a larger gluon distribution in the region from $x \lapproxeq 10^{-3}$ down to $10^{-5}$, at low scales, than coming from {\it extrapolations} of the NLO gluon PDFs from global fits to data not including the $\J$ data.
 An indication in favour of a larger gluon PDF in this domain comes also from the recent LHCb data on open charm (and beauty) \cite{LHCbg}.

Finally, it is useful to compare our approach with that of \cite{IPSW16}, 
where it was demonstrated that the re-summation of the BFKL-induced 
$(\alpha_S\ln(1/\xi))^n$ terms in the coefficient functions 
additionally reduces the factorization scale dependence.  Recall that
our choice of $\mu_F=M_\psi/2$ resums only the {\em double logarithmic},
$(\alpha_S\ln(1/\xi)\ln\mu_F)^n$ contributions\footnote{This result for the optimal scale (see Section \ref{sec:opt}) is confirmed by the formula after eq.~(8) in \cite{IPSW16}. Note that, in \cite{IPSW16}, 
$L(L-\ln 16)+\ln^24=(L-\ln 4)^2=\ln^2(M^2_\psi/4\mu^2_F)$, since $L\equiv {\rm ln}(M^2_\psi/\mu^2_F)$.}. The 
remaining 
part, which does not contain $\ln\mu_F$, should be considered, in the collinear 
factorization approach, as higher-order,
NNLO, N$^3$LO, ... corrections. Of course, it would be good to account for 
these corrections as well. However, to  properly calculate these corrections one has
to exclude the low ($<Q^2_0$) virtuality contribution. Otherwise we 
will face the problem of double counting again.  The present paper shows these (power) corrections (necessary to avoid double counting) are crucial to achieve perturbative stability.

\begin{figure} [t]
\begin{center}
\includegraphics[width=0.35\textwidth]{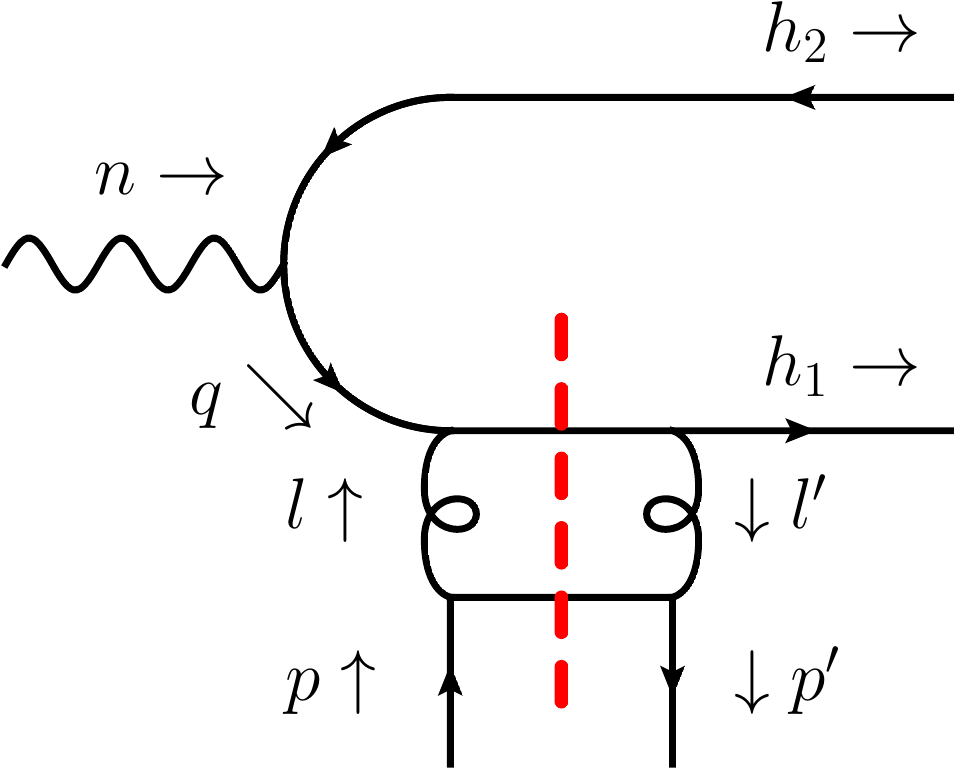}
\qquad
\includegraphics[width=0.35\textwidth]{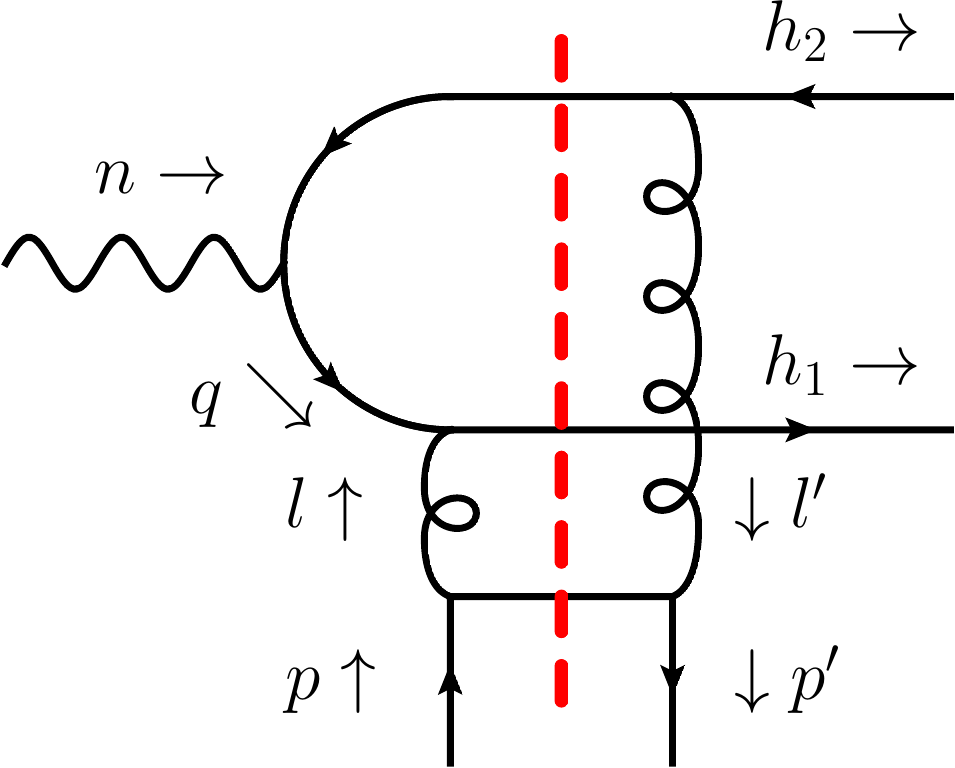}
\caption{\sf Two diagrams (a,b) computed for the NLO quark coefficient function. Note that $p$ and $p'$ refer to the incoming and outgoing quark lines. In the corresponding diagrams computed for the NLO gluon coefficient function  the light quark line is replaced by a gluon.  The other two diagrams of the different coupling of the two $t$-channel gluons to the heavy quarks are implicitly included.}
\label{fig:f4}
\end{center}
\end{figure}

\section*{Appendix}

Here we describe the calculation of the piece 
that we subtract from the full result. Only the imaginary part of the ladder-type cut diagrams shown in Fig.~\ref{fig:f4} and the corresponding diagrams where the light-quark line is replaced by gluons is computed.

All momenta appearing in the calculation may be decomposed in terms of light-like momenta $p, n$ and a transverse four-momentum $l_\perp$,
\begin{align}
& l^\mu = \beta p^\mu + \alpha n^\mu + l_\perp^\mu, &
&h_1^\mu = h_2^\mu = \beta_h p^\mu + \alpha_h n^\mu, &
\end{align}
where $l$ is the loop momentum and $h_1, h_2$ are the momenta of the outgoing heavy quark and heavy anti-quark, respectively. Here $p$ can be chosen as the momentum of the incoming light parton and $n$ the momentum of the incoming on-shell photon. With this convention we have
\begin{align}
& p \cdot p = n \cdot n = 0, & & p \cdot n = \hat{s}/2, & & p \cdot l_\perp = n \cdot l_\perp = 0, &
\end{align}
where $\hat{s}$ is the photon-parton centre-of-mass energy squared.
The four momenta of the incoming and the outgoing light partons are proportional. We may write $p_\mu$ and $ p'_\mu=Xp_\mu$ with  
\be
X=\frac{x-\xi}{x+\xi}=\frac{\hat{s}-M^2_{\psi}}{\hat{s}}=\frac{y}{1+y}, ~~~~~~{\rm where}~~~~~~ y=\frac{x-\xi}{2\xi}=\frac{\hat{s}}{M^2_\psi}-1.
\label{eq:X}
\ee

To leading order in the heavy quark relative velocity, the S-wave spin-triplet component of $\J$ can be computed using the projection \cite{Petrelli:1997ge,Bodwin:2002hg,Braaten:2002fi}
\begin{equation}
v_\alpha(h_2) \bar{u}_\beta(h_1) \rightarrow 
N_{\J}
\left[ (\slashed{h}_2 - m_c) \slashed{\epsilon}_{\J}^* (\slashed{K} + M_\psi) (\slashed{h}_1 +m_c) \right]_{\alpha \beta}.
\end{equation}
Here $\bar{u},v$ are the spinors of the outgoing heavy quark and anti-quark which form the $\J$. The indices $\alpha$ and $\beta$ label their spin. $N_{\J}$ is an overall factor which contains the non-perturbative NRQCD matrix element describing the $\J$ formation. The vector $\epsilon_{\J}$ describes the polarisation of the $\J$ with momentum $K=h_1+h_2$ and mass $M_\psi=2 m_c$.

The projections onto the quark and gluon GPDs are given by \cite{Ji:1996ek,Ji:1998xh,Radyushkin:1996ru,Radyushkin:1997ki},
\begin{align}
&u_\alpha(p) \bar{u}_\beta(p^\prime) \rightarrow N_q \slashed{p}_{\alpha \beta}~,&
&\epsilon_1^{\mu} {\epsilon^*_2}^{\nu} \rightarrow N_g g_\perp^{\mu \nu} = N_g \left( g^{\mu \nu} - \frac{2}{\hat{s}} p^\mu n^\nu - \frac{2}{\hat{s}} n^\mu p^\nu \right),&
\end{align}
respectively. Here $u, \bar{u}$ are the spinors of the light quarks connected to the quark GPD and $\epsilon_1$, $\epsilon^*_2$ are the polarisation vectors of gluons connected to the gluon GPD. $N_q$, $N_g$ are overall factors containing the quark and gluon GPDs.

The on-shell conditions $h_1^2 - m_c^2 = 0$ and $h_2^2 - m_c^2 = 0$ for outgoing heavy quarks and the cut-constraints, $(p-l)^2 = 0$ and $(n-h_2+l)^2 -m_c^2 = 0$ for Fig.~\ref{fig:f4} diagram (a), $(p-l)^2=0$ and $(h_1-l-n)^2 - m_c^2 = 0$ for Fig.~\ref{fig:f4} diagram (b), allow us to choose $\alpha_h = 1/2, \beta_h = 2 m^2/\hat{s}$ and fix $\alpha, \beta$ in terms of $l_\perp^2, \hat{s}, m_c$.  Specifically,
\begin{align}
&\beta=4(1+l^2/\hat{s})m_c^2/\hat{s}-2l^2/\hat{s},&
&\alpha=l^2/\hat{s},&  \label{eq:beta} \\
&l^2=l^2_\perp/(1-\beta),& 
&l^{'2}=l^2(1-4m_c^2/\hat{s}).&
\end{align}
Additionally, we obtain $q^2=-m_c^2$  for diagram \ref{fig:f4}(a) and
\be
q^2=l^2_\perp-\left(\frac{\hat{s}}{2}-l^2\right)\left(\beta-\frac{2m_c^2}{\hat{s}}\right)~=~3m_c^2-\beta \hat{s}
\ee
for diagram \ref{fig:f4}(b).



In our calculation we split each diagram of Fig.~\ref{fig:f4} into two parts. An ``upper'' part which contains a trace over the heavy quark fermion line and a ``lower'' part which in the quark channel contains a trace over the light quark line and in the gluon case consists of two triple gluon vertices contracted with $g_\perp^{\mu \nu}$.

First we discuss the ``upper'' part which is different for the diagrams (a) and (b) of Fig.~\ref{fig:f4} but identical for the quark and gluon channels.
Where it appears, we replace the contraction of $l_\perp$ with the polarisation vectors using
\begin{equation}
(l_\perp \cdot \epsilon^*_{\J}) (l_\perp \cdot \epsilon_\gamma) = (\epsilon_\gamma \cdot \epsilon^*_{\J}) l_\perp^2/2
\end{equation}
which follows from tensor decomposing the $l_\perp$ integral after the integration over the $\vec{l}$ azimuthal angle. We can simplify the calculation by noting that the sum of the ``upper'' parts of diagrams (a) and (b) obey the gauge condition
\begin{equation}
\mbox{T(h.loop)}^{\mu\nu}l_\mu=\mbox{T(h.loop)}^{\mu\nu}l'_\nu=0 , \label{eq:gaugecondition}
\end{equation}
where
\be
\mbox{T(h.loop)}^{\mu\nu} =  \frac{1}{(-2m_c^2)}\mbox{Tr(h.loop)}_a^{\mu\nu} + \frac{1}{(2 m_c^2 - \beta \hat{s})}\mbox{Tr(h.loop)}_b^{\mu\nu}.
\ee
Here T(h.loop) is the upper part of the amplitude, which besides the trace over the quark loop, includes the heavy quark propagator $1/(q^2-m^2_c)$.

Using the gauge condition the only contractions of the ``upper'' part that appear in the sum of diagrams are
\bea
\mbox{Tr(h.loop})^{\mu\nu}_ag_{\mu\nu} &=& N_{\J}\, 4m_c(\epsilon_\gamma \cdot \epsilon^*_{J/\psi})(6m_c^2-\hat{s}\beta),  \\
\mbox{Tr(h.loop})^{\mu\nu}_ap_\mu p_\nu &=& N_{\J}\, 4m_c(\epsilon_\gamma \cdot \epsilon^*_{J/\psi})\hat{s}^2(1/2+\alpha)/2,  \\
\mbox{Tr(h.loop})^{\mu\nu}_ap_\mu l_{\perp\nu} &=& N_{\J}\, 4m_c(\epsilon_\gamma \cdot \epsilon^*_{J/\psi})l^2_\perp \hat{s}/2,  \\
\mbox{Tr(h.loop})^{\mu\nu}_al_{\perp\mu} p_\nu &=& N_{\J}\, 4m_c(\epsilon_\gamma \cdot \epsilon^*_{J/\psi})l^2_\perp \hat{s}/2, \\
\mbox{Tr(h.loop})^{\mu\nu}_bg_{\mu\nu} &=& N_{\J}\, 4m_c(\epsilon_\gamma \cdot \epsilon^*_{J/\psi})2(\hat{s}\alpha\beta-m_c^2(2\alpha+1)), \\
\mbox{Tr(h.loop})^{\mu\nu}_bp_\mu p_\nu &=& - N_{\J}\, 4m_c(\epsilon_\gamma \cdot \epsilon^*_{J/\psi})\hat{s}^2/4, \\
\mbox{Tr(h.loop})^{\mu\nu}_bp_\mu l_{\perp\nu} &=& - N_{\J}\, 4m_c(\epsilon_\gamma \cdot \epsilon^*_{J/\psi})l^2_\perp(\hat{s}-l^2)/2,  \\
\mbox{Tr(h.loop})^{\mu\nu}_bl_{\perp\mu} p_\nu &=& N_{\J}\, 4m_c(\epsilon_\gamma \cdot \epsilon^*_{J/\psi})l^2_\perp (\hat{s}+l^2)/2.
\eea
The contractions involving $p^\mu n^\nu, n^\mu p^\nu, n^\mu n^\nu, n^\mu l_\perp^\nu, l_\perp^\mu n^\nu, l_\perp^\mu l_\perp^\nu$ appear in the computation of individual diagrams but cancel for the sum of diagrams.

\subsection*{Quark-induced NLO correction}

For an unpolarized light quark the trace over the ``lower'' light quark line gives
\be
A^q_{\mu\nu}= 4 N_q~[p_\mu(p-l)_\nu+(p-l)_\mu p_\nu +g_{\mu\nu}(p\cdot l)]\ ,
\label{a1}
\ee
where the normalization factor 
\begin{equation}
N_q= C_F F_q(x,\xi,\mu_F)
\end{equation} 
includes the colour factor $C_F$ and the quark GPD, $F_q$.

This light quark part should be contracted with the trace, ${\rm Tr(h.loop)}^{\mu \nu}$, given by the heavy quark (upper) loop. Due to the gauge condition \eq{gaugecondition} we have that $(p-l)_\mu$ acts as $p_\mu$, while $(p-l)_\nu$ acts as $p'_\nu=Xp_\nu$ giving
\begin{align}
M^q_a &= \frac{4 N_q}{(-2m_c^2) \, l^2 {l^\prime}^2} \left[ \mbox{Tr(h.loop})^{\mu\nu}_ag_{\mu\nu} \left( \frac{\alpha \hat{s}}{2} \right) + \mbox{Tr(h.loop})^{\mu\nu}_ap_\mu p_\nu (1 + X) \right] + \overline{M}^q \\
&= \frac{4 N_q N_{\J} (2m_c) (\epsilon_\gamma \cdot \epsilon^*_{J/\psi}) }{(-2m_c^2) \, l^2 {l^\prime}^2} \left[(6m_c^2-\hat{s}\beta) \alpha \hat{s} + \hat{s}^2(1/2+\alpha)(1+X) \right] + \overline{M}^q, \label{a2}
\end{align}
for diagram (a) and 
\begin{align}
M^q_b &= \frac{4 N_q}{(2 m_c^2 - \beta \hat{s}) \, l^2 {l^\prime}^2} \left[ \mbox{Tr(h.loop})^{\mu\nu}_bg_{\mu\nu} \left( \frac{\alpha \hat{s}}{2} \right) + \mbox{Tr(h.loop})^{\mu\nu}_bp_\mu p_\nu (1 + X) \right] - \overline{M}^q \\
&= \frac{4 N_q N_{\J} m_c (\epsilon_\gamma \cdot \epsilon^*_{J/\psi})}{(2 m_c^2 - \beta \hat{s}) \, l^2 {l^\prime}^2}\ [4 (\hat{s}\alpha\beta-m_c^2(2\alpha+1)) \alpha \hat{s} - (1+X) \hat{s}^2] - \overline{M}^q, \label{b2}
\end{align}
for diagram (b). The term $\overline{M}^q$ accounts for terms which cancel between the two diagrams. The denominators come from the uncut propagators: $1/l^2$ for the left, and $1/l^{'2}$ for the right gluon and $1/(q^2-m_c^2)$ for the uncut heavy quark propagator.  

The result is to be integrated over the gluon transverse momentum ($dl^2_\perp$) while the longitudinal components are fixed by the quark on-mass-shell conditions. It is easy to perform this integral numerically accounting for the condition  which was introduced in Section 2 in order to avoid double counting.
 Recall, however, that we are not going to calculate the whole NLO contribution, but just the correction to the known $\overline{\rm MS}$ coefficient function. So, in order to compute the {\em correction}, in the integration  over the $l_\perp$ we only consider the region of $|l^2|<Q^2_0$.
 Actually, we integrate over $dl^2$ directly; the factor $(1-\beta)$ coming from the relation $l^2=l^2_\perp/(1-\beta)$ is exactly cancelled by the residue from the light quark on-mass-shell pole. 
 So we obtain the correction to the quark-induced part of the $\gamma p\to \J+p$ amplitude
 \be
 \Delta{\rm Im}{\cal M}^q~=~\frac{\alpha_s^2}{2\pi}\int^1_\xi dx\left(F_q(x,\xi,m_c)-F_q(-x,\xi,m_c)\right)\left(\int^{Q_0^2}_0   (M^q_a+M^q_b) \frac{2\pi m^4_c}{{\hat s}^2}dl^2 \right)
 \ee
where the `hard matrix elements' $M^q_{a,b}$ are given by (\ref{a2}) and   (\ref{b2}). The factor $1/\hat{s}^2$ comes from the delta functions needed to put the lower light quark and the heavy quark coupled to the right gluon in Fig.~\ref{fig:f4} on-mass-shell.  The factor $m_c^4$ accounts for the normalization $N_{\J}$, defined
to be consistent with the normalization of eqs.~(3.93) and (3.95) of \cite{ISSK} for which the correction was calculated; actually the last factor (...) is the correction to $f_q$ of (3.93) of \cite{ISSK}.\footnote{The overall normalization has been checked against \cite{ISSK} and correctly reproduces the leading log term $\propto{\rm ln}(4m_c^2/\mu^2_F)$.}  For the gluon correction $\Delta{\cal M}^g$ there is an additional factor $\hat{s}/2m_c^2=1/\xi$ due to the definition of the gluon GPD, $F_g$;      see the extra factor of $\xi$ in eq. (3.94) of \cite{ISSK}, see also \cite{SJonesThesis}.


Note that we have explicitly calculated the NLO diagrams (a) and (b) of Fig.~\ref{fig:f4} which contain both LO~\footnote{The integration of the pure logarithmic form $dl^2/l^2$ up to $\mu_F$ actually reproduces the LO contribution already included in Fig.~\ref{fig:f2}(a). On the other hand some non-logarithmic corrections originating from higher powers of $l^2$, together with the whole contribution above $\mu_F$, are NLO $\alpha_s$ corrections which are not enhanced by the large collinear $(l^2)$ logarithms. }
and NLO contributions.
To identify the NLO part we therefore have to subtract the contribution generated by the LO evolution equation, which is of the form of the convolution $P^{\rm LO}\otimes C^{\rm LO}$, before we integrate over $l^2_\perp$. This subtraction completely cancels the logarithmic infrared divergence $dl^2/l^2$. Note that the subtraction must be done only in the region of $|l^2|<\mu^2_F$ since at the factorization scale $\mu_F$ the DGLAP evolution stops.\footnote{This is the origin of the $\ln(4m^2/\mu^2_F)$ factor 
in the first term of $f_q(y)$ of eq.~(3.93) of \cite{ISSK}. Since now we integrate over the $|l^2|<Q^2_0<\mu^2_F$ the correction does not depend on $\mu_F$. } Also note that in the LO approximation the convolution $P^{\rm LO}\otimes C^{\rm LO}$ is larger than the value of the matrix element given by explicit calculation of the diagrams shown in Fig.~\ref{fig:f4}. Thus the final result has the sign opposite to that for the LO amplitude.

In this way we obtain the quark NLO coefficient function. Since we are looking for the power correction needed to avoid double counting of the low $|l^2|<Q^2_0$
 contribution\footnote{This contribution is already included in 
the input value GPD($Q_0)$.}, we actually have to integrate the matrix element $M^q$ over $|l^2|<Q^2_0$ only (as explained above) and to subtract the result from the known NLO coefficient function given in the $\overline{\rm MS}$ scheme.

In the notation of Ref.~\cite{ISSK} this should be considered as the new form of Im$f_q(y)$ of their eq.~(3.93), after allowing for the changes made by our introduction of the `$Q_0$ cut'.

\subsection*{Gluon NLO correction}

In the gluon case the tensor $A^g_{\mu\nu}$ corresponding to the lower part of Fig.~\ref{fig:f4} diagrams (with the lower quark line replaced by a gluon line) was calculated explicitly. It can be written in the form
\be
 A^g_{\mu\nu}= N^g(ag_{\mu\nu}+b_{11}p_\mu p_\nu+b_{22}h_\mu h_\nu+b_{12}p_\mu h_\nu +b_{21}h_\mu p_\nu+c_1p_\mu l_{\perp\nu}+d_1l_{\perp\mu} p_\nu+c_2h_\mu l_{\perp\nu}+d_2l_{\perp\mu} h_\nu)\ ,
\label{ag}
\ee
where $h_\mu=p_\mu-l_\mu\,$ and 
\be
a=l^2(1+X+4(1-\beta)),\;\;\;\;\;\;\; b_{11}=X(4\beta-2)-4(1-\beta)\ ,
\ee
$$b_{22}=2,\ b_{12}=2X+4,\  b_{21}=2+4X,\ c_1=3-2X,\ d_1=3X-2,\ c_2=3,\ d_2=3\ .$$
Here the normalization factor is \footnote{Here the denominator $(x+\xi-i\epsilon)(x-\xi+i\epsilon) $ arises from the particular definition of the gluon GPD.}
\be
N_g~=~\frac{C_A}{8} \frac{F_g(x,\xi,\mu_F)}{(x+\xi-i\epsilon)(x-\xi+i\epsilon)}.
\ee
Note that $X$ is defined in (\ref{eq:X}) and $\beta$ is given by (\ref{eq:beta}).  Recall that we are looking for the imaginary part of the amplitude (i.e. $s$-channel  discontinuity).

This expression should be convoluted with the ``upper'' part of the diagram.
The result for the sum of diagrams can again be simplified using the gauge conditions \eq{gaugecondition}. That is vector $h_\mu=(p-l)_\mu$ acts as $p_\mu$, while $h_\nu$ acts as $p'_\nu=Xp_\nu$.

As before, the result is multiplied by the terms $1/l^2$ and $1/l^{'2}$ from the $t$-channel gluon propagators  and by the term $1/(q^2-m_c^2)$ from the corresponding heavy quark propagator. Then we have to subtract the part generated by the LO evolution equation which is given by the convolution $P^{\rm LO}\otimes C^{\rm LO}$. Finally we integrate over $l^2_\perp$, accounting for the
 condition $|l^2|<Q^2_0$  (the longitudinal components are fixed by the 
heavy quark and gluon $(p-l)^2=0$ on mass-shell conditions). In this way we obtain the power correction which should be subtracted from the known NLO gluon coefficient function ${\rm Im}f_g(y)$ given by eq.~(3.95) of \cite{ISSK} (see also \cite{SJonesThesis}), which we then use to obtain the $Q_0$ subtracted NLO gluon contribution.

\section*{Acknowledgements}

We thank Ronan McNulty for interesting discussions and for encouraging
us to make these predictions. MGR thanks the IPPP at the University of 
Durham for hospitality. MGR is supported by the RSCF grant 14-02-00281.  SPJ is supported by the
Research Executive Agency (REA) of the European Union under the Grant
Agreement PITN-GA2012316704 (HiggsTools), and TT is supported by STFC 
under the consolidated grant ST/L000431/1.

\end{document}